# Whole-Building Fault Detection: A Scalable Approach Using Spectral Methods

**Michael Georgescu, PhD**     **Sophie Loire, PhD**     **Don Kasper**     **Igor Mezic, PhD**
*Affiliate Member ASHRAE*

## ABSTRACT

*In this paper, an extension to rules-based fault detection is demonstrated utilizing properties of the Koopman operator. The Koopman operator is an infinite-dimensional, linear operator that captures nonlinear, finite dimensional dynamics. The definition of the Koopman operator enables algorithms that can evaluate the magnitude and coincidence of time-series data. Using spectral properties of this operator, diagnostic rule signals generated from building management system (BMS) trend data can be decomposed into components that allow the capture of device behavior at varying time-scales and to a granular level. As it relates to the implementation of fault detection (FDD), this approach creates additional spatial and temporal characterizations of rule signals providing additional data structure and increasing effectiveness with which classification techniques can be applied to the analysis process. The approach permits a knowledge base to be applied in a similar manner to that of a rules-based approach, but the introduced extensions also facilitate the definition of new kinds of diagnostics and overall provide increased analysis potential.*

## INTRODUCTION

Commercial buildings substantially impact national energy use as they represent 18% of the total primary energy consumption in the United States (DOE 2009). Of the processes that constitute this energy usage, a significant opportunity to reduce consumption lies in the efficient operation of HVAC systems. Poorly maintained and improperly controlled HVAC equipment is responsible for 15% to 30% of the energy use of a building's HVAC system (Basarkar 2013). In addition to loss of energy efficiency, added operational costs are incurred from decreased productivity from reduced occupant comfort and increased maintenance (Fowler 2011).

In previous studies examining the efficiency of building HVAC operation, the effects causing the observed variance between design expectations and actual performance has primarily been attributed to operational problems such as improper installation, equipment degradation, sensor failures, or control logic problems (Djuric & Novakovic 2009, Wang et al. 2013). Numerous studies highlighting examples of building underperformance can be found in literature. In particular, the work of (Torcellini et al. 2006) contains several prominently referenced case studies. Building underperformance is not limited to buildings that are new or old, or buildings designed with energy efficiency in mind. A conclusion echoed throughout industrial case studies is that underperformance is caused by insufficient monitoring where the lack of availability of building operational data is a key determinant of success in this realm.

Although fundamentally buildings exist to provide a comfortable environment for occupants, a barrier to effective building monitoring is that no two buildings have the same system setup, physical location, or occupancy

**Michael Georgescu** is the Director of Engineering & Research at Ecorithm Inc., Santa Barbara, California. **Don Kasper** is the Vice President of Operations at Ecorithm Inc., Santa Barbara, California. **Sophie Loire** is a Research Fellow & Technical Fellow at Ecorithm Inc., Santa Barbara, California. **Igor Mezic** is a professor in the Department of Mechcanical Engineering, University of California, Santa Barbara, California and a co-founder of Ecorithm Inc.

patterns. Two substantial challenges are posed in the analysis of HVAC systems: the first is the task of automatically analyzing data from a complex building, and the second is the obstacle of repeating effective analysis at scale. For more information on the different approaches to building diagnostics, refer to the review paper of (Katipamula 2005).

Beginning with the adoption of computerized building management systems, the HVAC industry has seen numerous solutions aimed at improving monitoring efficiency and augmenting the capabilities of facility managers. Commercially, solutions for the fault detection and diagnostics of HVAC systems have gained traction in industry. Many of these implementations conceptually stem from the work of (House et al 2001, Schien & House 2003) on rules-based diagnostics which have gained adoption due to ease of implementation. Alternative model-based approaches also exist, but implementation has been limited. A review paper comparing analysis toolkits in (Lee et al 2015) contains an overview of data-based and model-based approaches. Although these techniques are automatable, challenges still exist in application at scale due to the heterogeneous nature of building system configurations.

In this work, a technique to extend the analysis of the rules-based FDD approach is introduced utilizing spectral properties of the Koopman operator. By decomposing the output of a rules-based FDD implementation into modes that capture the behavior of rule error dynamics at varying time-scales, the magnitude and coincidence between rules can be evaluated. In this view, the behavior of rules can be differentiated at a more granular level which enable the detection of issues which occur on a shorter duty cycle, the definition of more generalized rule-sets, and thus a more scalable analysis approach.

The remainder of this paper is organized as follows: in the following section, a brief overview of the Koopman operator is given. This is followed by a series of examples demonstrating the advantages of this approach. The paper is concluded with a discussion on how this approach can extend data-driven fault detection.

**THE KOOPMAN OPERATOR**

In the analysis of buildings, as with any other dynamical system, understanding system behavior is critical for performing analysis. Because the equations describing the behavior of a building are of high dimension, and impractical to express analytically, data-driven methods are invaluable in the analysis of these systems. Properties of the Koopman operator are well suited for this context. By projecting the time-series of sensor measurements onto eigenfunctions of the operator, features of a system can be extracted.

To introduce the Koopman operator, consider the evolution of a nonlinear dynamical system given by:

$$x(t + \Delta t) = F(x(t)) \qquad (1)$$

where $x \in M$ are variables belonging to a finite, but multi-dimensional space $M$, and $F: M \to M$ maps the state variables at time $t$ to time $t + \Delta t$. The Koopman operator $U$ is a linear operator that acts on $M$ in the following manner: for $g: M \to \mathbb{R}$, where $g$ is a function describing observations of the state variables, $U$ maps $g$ to a new function $Ug$ given by

$$Ug(x) = g(F(x(t))) = g(x(t + \Delta t)) \qquad (2)$$

The Koopman operator describes the evolution of an observable one step in time, and iterative application of the operator describes the trajectory of the observable. The eigenfunctions and eigenvalues of the operator are defined as follows: for eigenfunctions $\Psi_k: M \to \mathbb{C}$ and constant eigenvalues $\lambda_k \in \mathbb{C}$

$$U\Psi_k(x) = \lambda_k \Psi_k(x) \qquad k = 1, \dots, n. \qquad (3)$$

It follows that vector-valued observables, $G: M \to \mathbb{R}^m$, that are in the span of eigenfunctions can be expressed by

$$G(x) = \sum_{k=1}^{n} \lambda_k \Psi_k(x) v_k. \tag{4}$$

In Eq. (4), $\{v_k\}_{k=1}^{n}$ are a set of vectors called Koopman modes (KM), and are coefficients of the projection of observables onto the eigenfunctions of the operator. Koopman modes describe the dynamics of observables at different time-scales. For dynamics on an attractor, Koopman eigenvalues are on the unit circle. This allows $G(x)$ to be computed by calculating Fourier averages, $G_\omega^*(x) \in \mathbb{C}^m$ of the form:

$$G_\omega^*(x) = \lim_{n \to \infty} \frac{1}{n} \sum_{j=0}^{n-1} e^{i2\pi j \omega} G(x(j)). \tag{5}$$

By applying $G_\omega^*(x)$ to the definition of the operator, it becomes clear that Fourier averages satisfy the eigenvalue equation for the operator.

The notion of Koopman modes was introduced in (Mezic 2005), and the connection between Fourier analysis and the Koopman operator was first described in (Mezic & Banaszuk 2004). There are several methods available for calculating Koopman modes such as taking Fourier averages over the spatial field, using the Arnoldi algorithm (Rowley et al. 2009), or Fourier transformation when observables are periodic (Susuki 2009). In contrast to Fourier analysis which is well suited to periodic signals, methods exist for calculating Koopman modes in cases where signals may be non-periodic, and non-uniformly spaced in time. For more information about modal decompositions using the Koopman operator, refer to the references above and the review paper of (Budišić et al. 2012).

## EXTENDING RULES-BASED ANALYSIS WITH SPECTRAL METHODS

The Koopman operator connects concepts in dynamical systems theory with Fourier analysis. A notion to be aware of is that the analysis of data is performed in the frequency domain (i.e. on-attractor dynamics whose eigenvalues are on the unit circle) as opposed to conventional analysis performed in the time-domain. In the frequency domain, complex values are associated with dynamics at varying time-periods which are derived from an entire diagnostic signal. This provides a characterization of the overall behavior of signal.

In the figure below, an example diagnostic signal is shown in both the time-domain and frequency domain.

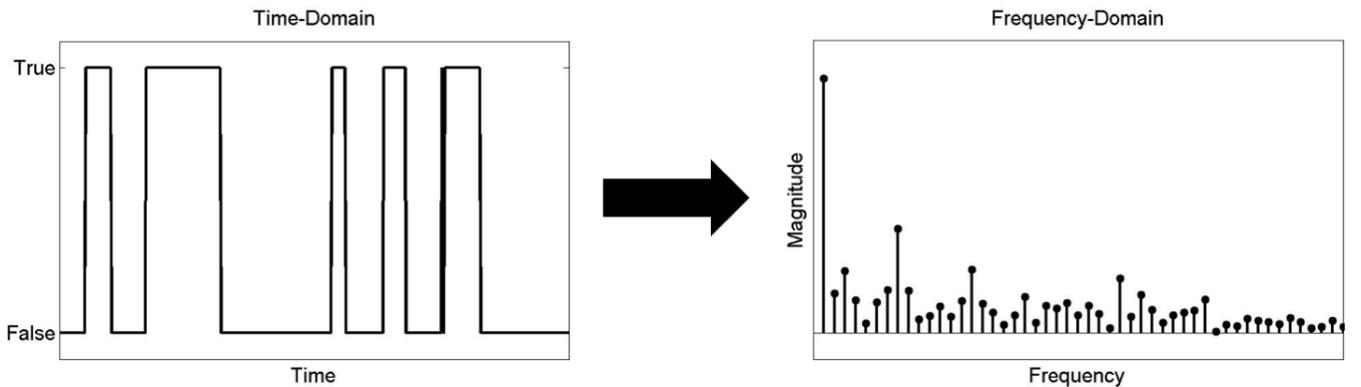

**Figure 1**   Representations of a diagnostic rule signal: time-domain (left) and frequency domain (right)

In the time-domain, the diagnostic signal is a series of 1's and 0's indicating whether the diagnostic condition is true or false. In the frequency-domain, the same diagnostic signal is represented by a series of complex values. The complex values correspond to magnitudes and phases which are associated with their respective time-periods. In this representation, one can examine the time-scales at which the diagnostic signal is true instead of individual measurements on a per time-step basis. The advantage of the frequency-domain representation is that global characteristics of the diagnostic signal can be easily evaluated in comparison to the time-domain view.

## Characterizing Time-Scales

Let $G(x)$ be the time-series generated by applying a diagnostic rule to building data. In the figure below, the values in time of three time-series diagnostic signals are shown. It is clear that these time-series signals have qualitatively different behaviors in time, however, from a statistical point of view, the three signals have similar average values (i.e. percentage of time the rule is violated) and standard deviation. In this case, standard statistical measures do a poor job of differentiating the behavior of these signals from one another.

**Table 1. Statistical Properties of Diagnostic Signals**

| Statistic | Signal 1 (S1) | Signal 2 (S2) | Signal 3 (S3) |
|---|---|---|---|
| Average | 0.41 | 0.54 | 0.43 |
| Standard Deviation | 0.49 | 0.50 | 0.50 |

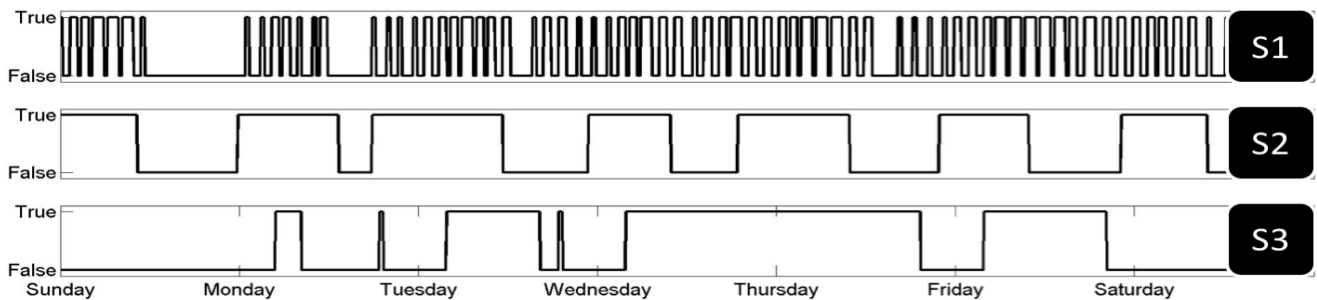

**Figure 2** Three diagnostic rule signals. Each signal exhibits different oscillatory characteristics.

Calculating the Koopman modes of these three signals generates a frequency-domain representation of the data. In the figure below, the spectral response of the three diagnostic signals is shown.

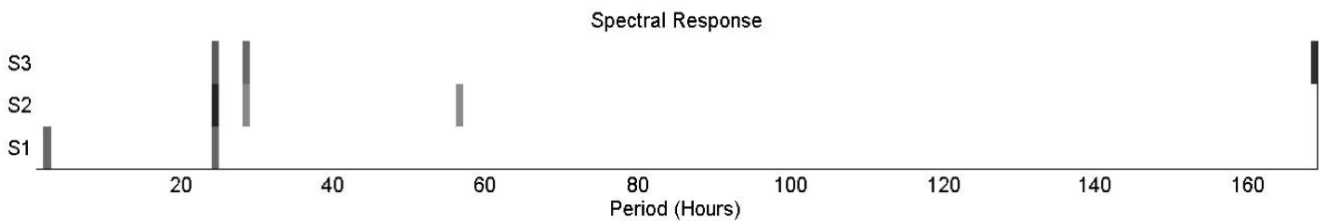

**Figure 3** Spectral response of signals 1,2, and 3 (previously shown in Figure 2).

Figure 3 visualizes the magnitude of the most significant Koopman mode elements to demonstrate time-scales of value change in the diagnostic signals. It is clear from the visualization that signal 3 contains most of its spectral energy at the 168 hour (weekly) period, signal 2 contains most of its spectral energy at the 24 hour (daily) period, and signal 1 contains most of its spectral energy at the 2 hour period. Compared to analyzing statistical properties of these signals, this representation allows for the differentiation of signal behaviors. The additional resolution can help to better isolate the root causes of issues. Faults can occur in a variety of ways; static faults are continuous and often relate to a hardware malfunction. Alternatively, design limitations in equipment often reveal themselves in a degradative pattern occurring over long oscillatory scales (e.g. multiple days or over a week), and programmatic/controller issues can often occur abruptly by exhibiting high frequency oscillations in the activation of associated diagnostic signals. These characteristics are more difficult to extract using only statistical properties of diagnostic signals which don't capture signal frequencies.

## Spatial Representations

Now the study of Koopman modes will be extended to analyze behavior in a spatial context. Let $\{g_i\}_{i=1}^m = G(x)$ be the time-series generated by applying a diagnostic rule to building data where each observable function $g_i$ is associated with a physical zone of the building. In this case, the i-th element of KM $v_k$ is the i-th zone's spectral energy corresponding to the k-th KM. In the figure below, the magnitude and angle of the 24 hour KM of a set of diagnostic signals is plotted spatially on a view of a building floorplan.

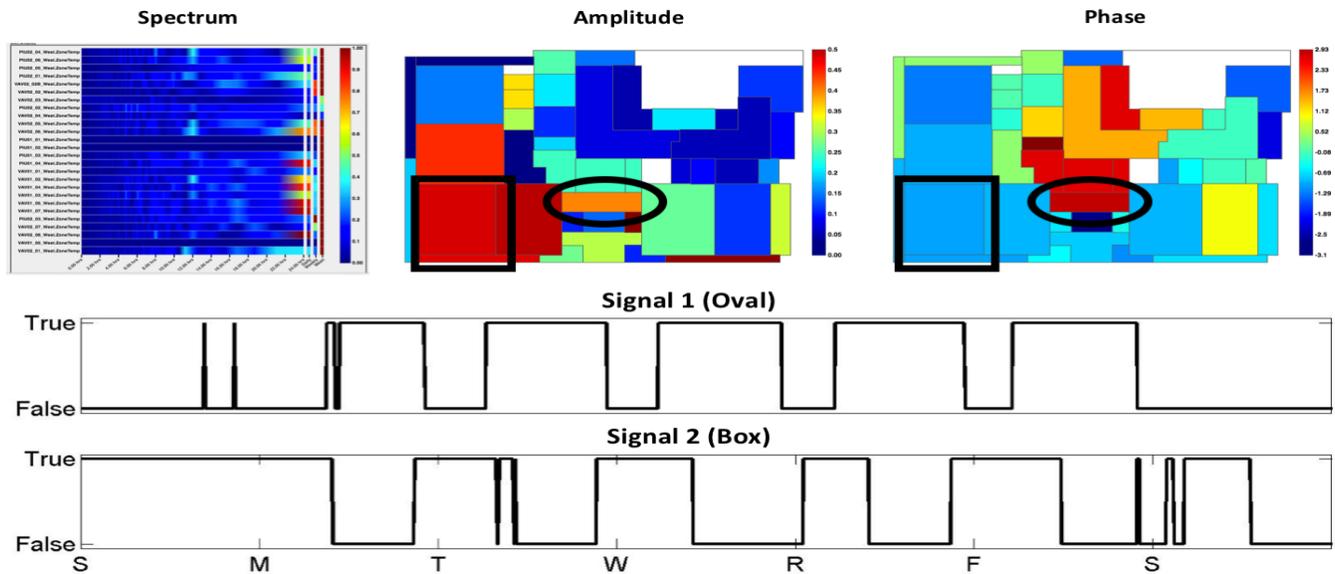

**Figure 4**   (A) spectrum, amplitude and phase of 24hr spectral response plotted on building plan view (top), (B) value vs time of two zones (bottom) denoted by box and oval on building plan view.

In Figure 4, two zones (represented by signal 1 and signal 2) are denoted which have different values of KM magnitude and phase. Their associated time-series are shown for reference. These zones have a difference in phase of almost π radians that is reflected in the associated time-series. The diagnostic signals for these two zones are non-coincident at the 24 hour time scale since they are out of phase. Additionally, the magnitude for zone 1 is greater than that for zone 2. This is also reflected in the time-series as the amount of time, on a daily time-scale, that the diagnostic signal is true is greater for zone 1 than it is for zone 2.

By associating the values of Koopman modes with associated zones, the magnitude and coincidence of diagnostic signals can be evaluated and aggregated along different spatial dimensions. Whether it is a plan view of a building illustrating adjacency between spaces as shown in Figure 4, or a hierarchical view showing the physical connectivity of HVAC devices, the spatial representation allows the creation of additional groupings of diagnostic signals to help identify issues that are systemic in nature. When represented as a graph, these spatial characteristics can be utilized algorithmically allowing one to further associate issues with physical building elements (e.g. specific floors, building orientations, or adjacencies between a subset of rooms).

## Information Compression

Since most of the spectral energy of a diagnostic signal is limited to only a few frequencies, analysis can be restricted to only these frequencies to examine the bulk trend of a signal's behavior. In the work of (Georgescu 2013, Georgescu 2014, Georgescu 2015) the reasoning for selecting various frequencies (168hr, 24hr, 12hr, 8hr, …) is discussed as it relates to different physical and informational contexts in building analysis. Because many factors in the behavior and operation of buildings are cyclical in time, from weather patterns that occur on daily and seasonal time-scales (Eisenhower 2010) to occupancy and operational patterns which occur at weekly, daily, and hourly time scales, the overall behavior of a diagnostic signal can often be described by examining a subset of frequencies. The figure below shows two diagnostic rule signals. The first signal illustrates a sustained rule violation while the second signal shows a rule that is violated in bursts, but over a similar overall time. Both signals are compared to an approximation generated by Eq. 4 where Koopman modes from only 3 frequencies (mean, 24hr, and 12hr periods) are preserved.

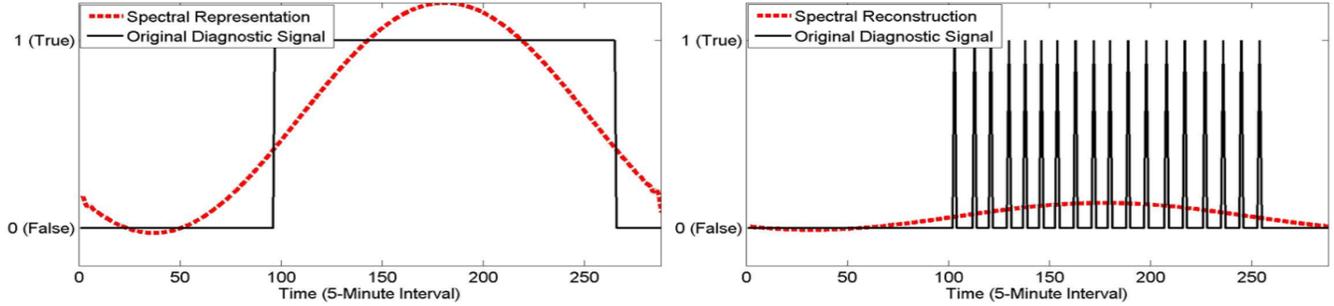

**Figure 5** (a) Rule signal true for a sustained period (left) (b) Rule signal which is true in bursts (left)

In Figure 5 the Koopman approximation does an excellent job representing the global characteristics of each signal. Since both signals are true over a similar overall duration, the Koopman approximation of these two signals have a similar phase. Because the first signal is true for a sustained period while the second signal is only true in bursts, the signal with a sustained rule violation has a larger amplitude in its spectral reconstruction. The advantage this provides in analyzing building data is that it enables the overall characteristics of diagnostic signals to be represented using fewer variables. In the original diagnostic signal, 2016 variables are required to fully represent it where with KMs, only three frequencies are needed to capture the overall characteristics. In Figure 5, the signals are dissimilar in a point-by-point comparison, by selecting time-scales on a larger time-scale, global characteristics of the signal can be compared. By selecting a subset of frequencies, signals can be compared by the time-scales that are present in the signal response.

**Rule Comparison**

In the previous sections, properties of the Koopman operator were demonstrated to illustrate how an error signal can be decomposed and evaluated across varying temporal and spatial scales using a reduced set of variables. Merging these aspects together, the behavior of multiple objects (i.e. different rules spanning single or multiple devices) can be compared.

Given a set of rule-signals $\{g_i\}_{i=1}^m = G(x)$, the similarity between signals can be evaluated by taking the inner product between KM elements. Based on the evaluated frequencies, the analysis resolution is modified. From these modes, the coincidence, $c_{x,y}$, between signals may be calculated through the following product:

$$c_{x,y} = \frac{\sum_{k=1}^n <v_{y,k.},v_{x,k.}>}{\sum_{k=1}^n \|v_{y,k}\|\|v_{x,k.}\|}. \qquad (6)$$

Where $<*,*>$ and $\| * \|$ are vector inner product and norm respectivley. Figure 6 below illustrates the coincidence between a set of 25 diagnostic signals calculated over the mean, 24hr, and 12hr KM frequencies.

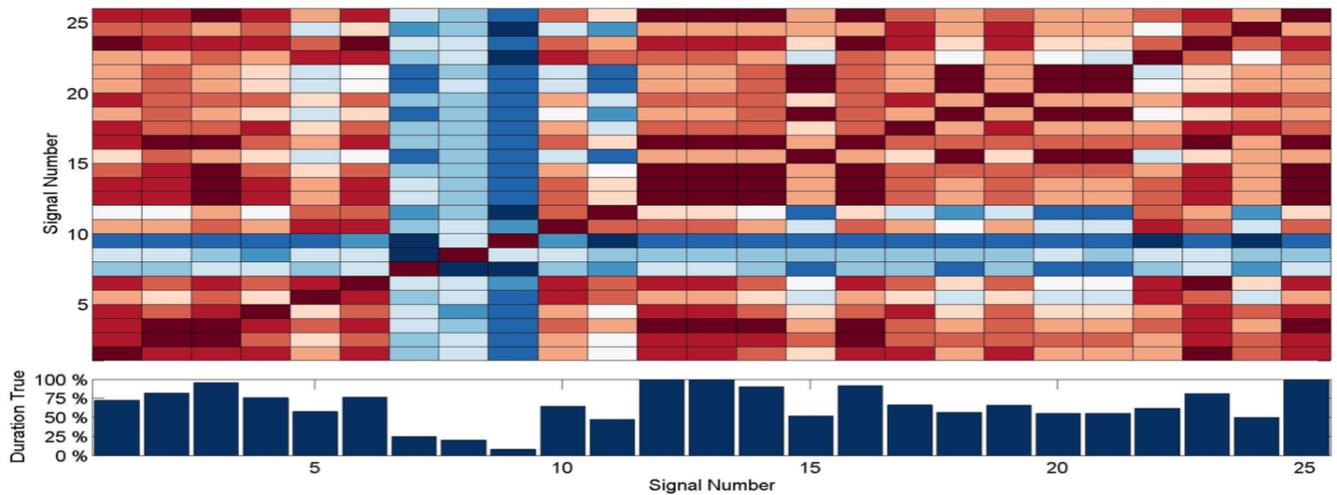

**Figure 6**  Coincidence matrix of 25 diagnostic signals. Red cells indicate diagnostic signal pairs with high similarity while blue cells indicate signal pairs with low similarity.

In Figure 6, the coincidence in response of different diagnostic signals are visualized relative to one another. Each element corresponds to the calculation of eq. (6). Indices in red indicate that a pair of signals are coincident with each other while indices in white or blue indicate that there is little coincidence between the given pair of signals. Using graphs representing spatial relationships between diagnostics, causal behaviors are easier to identify. Utilizing this representation of diagnostic signal data, combined with machine learning techniques, more granular diagnostic conditions may be defined and evaluated where faults are then expressed as an agglomerative combination of granular conditions. As traditional approaches scale, the number of diagnostic can grow rapidly as new diagnostics are defined which have partial overlap with existing diagnostics. The presented approach as only granular diagnostics are defined which can then be evaluated against each other based on their signal's magnitude and coincidence.

## DISCUSSION AND CONCLUSIONS

In this work, an approach for the analysis of output from diagnostics of building HVAC operation is introduced utilizing spectral properties of the Koopman operator. The method extends the capabilities of rules-based FDD approaches by decomposing the output of a diagnostic rule set applied to building HVAC data into varying temporal and spatial scales. The result is an analysis that isolates more granular behavioral properties of diagnostic signals. Because diagnostics are defined which can then be evaluated against each other based on their signal's magnitude and coincidence, fewer diagnostics are required to capture system behaviors. Additionally, there are fewer redundancies in the output generated. Fault classifiers designed under this approach have an increased ability to observe distinct system behaviors as the dimensionality of data is reduced from the presence of fewer diagnostics. The agglomerative nature of rule comparison enables is well suited for machine learning and manifold learning techniques which excel in settings where large amounts of structured data with little informational overlap exists. This paper focuses on simple rules to relay concepts but is rule agnostic.

## ACKNOWLEDGMENTS

The authors would like to acknowledge Thibaud Nesztler, Bradford Littooy, and Chris Tagge, for their contributions.

## NOMENCLATURE

$F$ = nonlinear function

| | | |
|---|---|---|
| $x$ | = | state variable |
| $M$ | = | high dimension manifold |
| $\mathbb{R}$ | = | set of real numbers |
| $\mathbb{C}$ | = | set of complex numbers |
| $U$ | = | Koopman operator |
| $g$ | = | observation function |
| $G$ | = | vector-valued observation function |
| $\Psi_k$ | = | k-th Koopman eigenfunction |
| $\lambda_k$ | = | k-th Koopman eigenvalue |
| $v_k$ | = | k-th Koopman mode of observable |

**Subscripts**

| | | |
|---|---|---|
| $k$ | = | eigenvalue / eigenfunction index |
| $m$ | = | dimension of state |